\documentclass[10pt,conference]{IEEEtran}
\usepackage[T1]{fontenc}
\usepackage{times}
\usepackage{amsmath}
\usepackage{amssymb}
\usepackage{booktabs}
\usepackage{array}
\usepackage{multirow}
\usepackage{graphicx}
\usepackage{hyperref}
\usepackage{xcolor}
\usepackage{cite}

\usepackage[top=54pt, bottom=54pt, left=44.65pt, right=44.65pt,
            columnsep=18pt]{geometry}

\begin{document}


\title{Scaling Photonic Tensor Cores with Unary \\and Homodyne Designs}

\author{
  \IEEEauthorblockN{Oluwaseun\ A.\ Alo and Ishan\ G.\ Thakkar}
  \IEEEauthorblockA{Department of Electrical and Computer Engineering, University of Kentucky, Lexington, KY, USA\\
  \{seun.alo, igthakkar\}@uky.edu}
}

\maketitle

\begin{abstract}
We analyze five photonic microring tensor core designs with a common optical power model. The results show that circuit ordering, unary encoding, and homodyne accumulation shape scalability, with the last two offering the strongest path to higher parallelism.
\end{abstract}
\begin{IEEEkeywords}
Photonic Tensor Cores, Scalability Analysis, Unary Encoding, Homodyne Superposition
\end{IEEEkeywords}

\section{Introduction}

Photonic microring tensor cores are promising for neural network inference because wavelength-division multiplexing offers much higher throughput and energy efficiency than electronic accelerators.\cite{holylight,deapcnn,sconna,astra,heana} Their basic engine is a vector dot-product core that performs the matrix multiplications in convolutional, fully connected, and attention layers.

Each core comprises a laser, a splitter, modulation, weighting, and optical summation at the photodetector. The block order defines the circuit organization: MAW (Modulation--Aggregation--Weighting), AMW (Aggregation--Modulation--Weighting), or MWA (Modulation--Weighting--Aggregation).\cite{holylight,deapcnn,spoga,heana} Because accumulated insertion loss grows along this path, organization directly limits scalable fan-in.

Data encoding then controls how precision trades against parallelism. Prior work uses three schemes: analog multi-level amplitude encoding,\cite{holylight,deapcnn,spoga} stochastic binary time encoding with bitwise logic,\cite{sconna,astra} and unary hybrid time–amplitude encoding, where pulse width and amplitude jointly encode input and weight.\cite{heana} In stochastic and unary designs, modulation and weighting merge into a single device, so any organization that aggregates between them (MAW) is impossible. The naive nine organization–encoding combinations therefore shrink to five realizable configurations.

A third axis, the superposition mode, sets the effective fan-in. Heterodyne designs sum many wavelengths at the photodetector and are bounded by free-spectral-range and inter-channel crosstalk.\cite{holylight,deapcnn,sconna,heana} Homodyne designs instead accumulate signals at one wavelength, removing inter-wavelength crosstalk and allowing fan-in to grow into the thousands.\cite{astra,spoga}

Existing accelerators each choose one configuration, but there is no unified scalability view of this design space. This work provides such a view by (i) enumerating the five valid configurations, (ii) applying a single optical power budget model to each, and (iii) quantifying spatial parallelism and temporal multiply–accumulate throughput, exposing the key trade-offs that should drive tensor core design.

\section{Organizations, Encodings, and Power Limits}

Fig.~1 summarizes the five realizable photonic tensor core configurations. The top row shows the three amplitude analog organizations. In MAW cores (HolyLight~\cite{holylight}) (Fig. 1(a)), modulation precedes wavelength aggregation, followed by a weight bank. In AMW cores (DEAPCNN~\cite{deapcnn}) (Fig. 1(c)), all wavelengths are first aggregated and split, then each branch applies modulation and weighting in series. In MWA cores (SPOGA~\cite{spoga}) (Fig. 1(b)), modulation and weighting occur before aggregation at the photodetector. All three configurations use heterodyne superposition at balanced photodetectors for summation of optical products.

The bottom row highlights the two non-amplitude-analog configurations, which exploit unary or time analog encoding to relax amplitude analog constraints and improve the scaling. In time analog MWA cores (HEANA~\cite{heana}), each time–amplitude optical modulator encodes input as pulse width and weight as pulse amplitude using a single microring. Multiple such modulators' outputs undergo incoherent superposition in balanced photo-charge accumulators (BPCAs) that provides in-situ spatial and temporal accumulation for a large number of products ~\cite{heana}. In unary homodyne MWA cores (ASTRA~\cite{astra}) (Fig. 1(d)), a single wavelength is shared among many optical stochastic multipliers whose outputs are combined by homodyne superposition and accumulated in PCAs, eliminating inter-wavelength crosstalk and removing the free-spectral-range defined wavelength parallelism ceiling ~\cite{astra}. Unary heterodyne AMW cores (SCONNA~\cite{sconna}) (Fig. 1(e)) cascade optical stochastic multipliers on different wavelengths after aggregation, using unary encoding to achieve high precision without relying on multi-level analog amplitudes~\cite{sconna}.

\begingroup
\scriptsize

\begin{equation}
B = \frac{1}{6.02}\left[20\log_{10}\!\left(\frac{R P_{\mathrm{PD\text{-}opt}}}
{\beta\sqrt{DR/\sqrt{2}}}\right) - 1.76\right]
\end{equation}

\begin{equation}
\begin{aligned}
\beta &= \sqrt{2q(RP_{\mathrm{PD\text{-}opt}}+I_d)
+ \frac{4kT}{R_L}
+ R^2 P_{\mathrm{PD\text{-}opt}}^2 \cdot RIN} \\
&\quad + \sqrt{2qI_d + \frac{4kT}{R_L}}.
\end{aligned}
\end{equation}

\begin{equation}\label{eq:pout}
\begin{aligned}
P_{\mathrm{out}} &= P_{\mathrm{Laser}} - P_{\mathrm{EC\text{-}IL}}
- P_{\mathrm{Si\text{-}att}} N d_{\mathrm{MRR}} \\
&\quad - P_{\mathrm{MRM\text{-}IL}} - (N{-}1)P_{\mathrm{MRM\text{-}OBL}} \\
&\quad - P_{\mathrm{split}}\log_2 M - P_{\mathrm{MRR\text{-}W\text{-}IL}} \\
&\quad - (N{-}1)P_{\mathrm{MRR\text{-}W\text{-}OBL}}
- P_{\mathrm{penalty}} - 10\log_{10}(N).
\end{aligned}
\end{equation}

\endgroup

\subsection{Scalability Model}

Scalability for all configurations is ultimately constrained by the optical power budget. The signal at each photodetector must exceed a minimum detectable power $P_{\mathrm{PD\text{-}opt}}$ that depends on the target precision $B$ and data rate $DR$ in Eqs (1) and (2). Given $P_{\mathrm{PD\text{-}opt}}$, the power arriving at the photodetector is~\cite{heana} given by Eq. (3). Here, $P_{\mathrm{penalty}}$ captures organization-specific losses and crosstalk penalties due to block ordering and layout. Reported values are: AMW: $5.8$\,dB, MAW: $4.8$\,dB, and MWA (analog or unary): $1.8$\,dB~\cite{heana,spoga}. 
For unary cores (ASTRA and SCONNA), the optical resolution is fixed at one bit in Eq. (1), which decouples numerical precision from analog dynamic range; precision scales instead with time or bit-stream length. 
We use device parameters from~\cite{heana}: $P_{\mathrm{Laser}}{=}10$\,dBm, $R{=}1.2$\,A/W, $I_d{=}35$\,nA, $R_L{=}50\,\Omega$, $\text{RIN}{=}{-}140$\,dB/Hz, $P_{\mathrm{EC\text{-}IL}}{=}1.6$\,dB, $P_{\mathrm{Si\text{-}att}}{=}0.3$\,dB/mm, $d_{\mathrm{MRR}}{=}0.02$\,mm, and $P_{\mathrm{MRM\text{-}IL}}{=}4$\,dB. We consider the free-spectral-range and comb lines limited multiplexing of a maximum number of 90 wavelengths per core.   

\begin{figure}[t]
\centering
\includegraphics[width=\columnwidth]{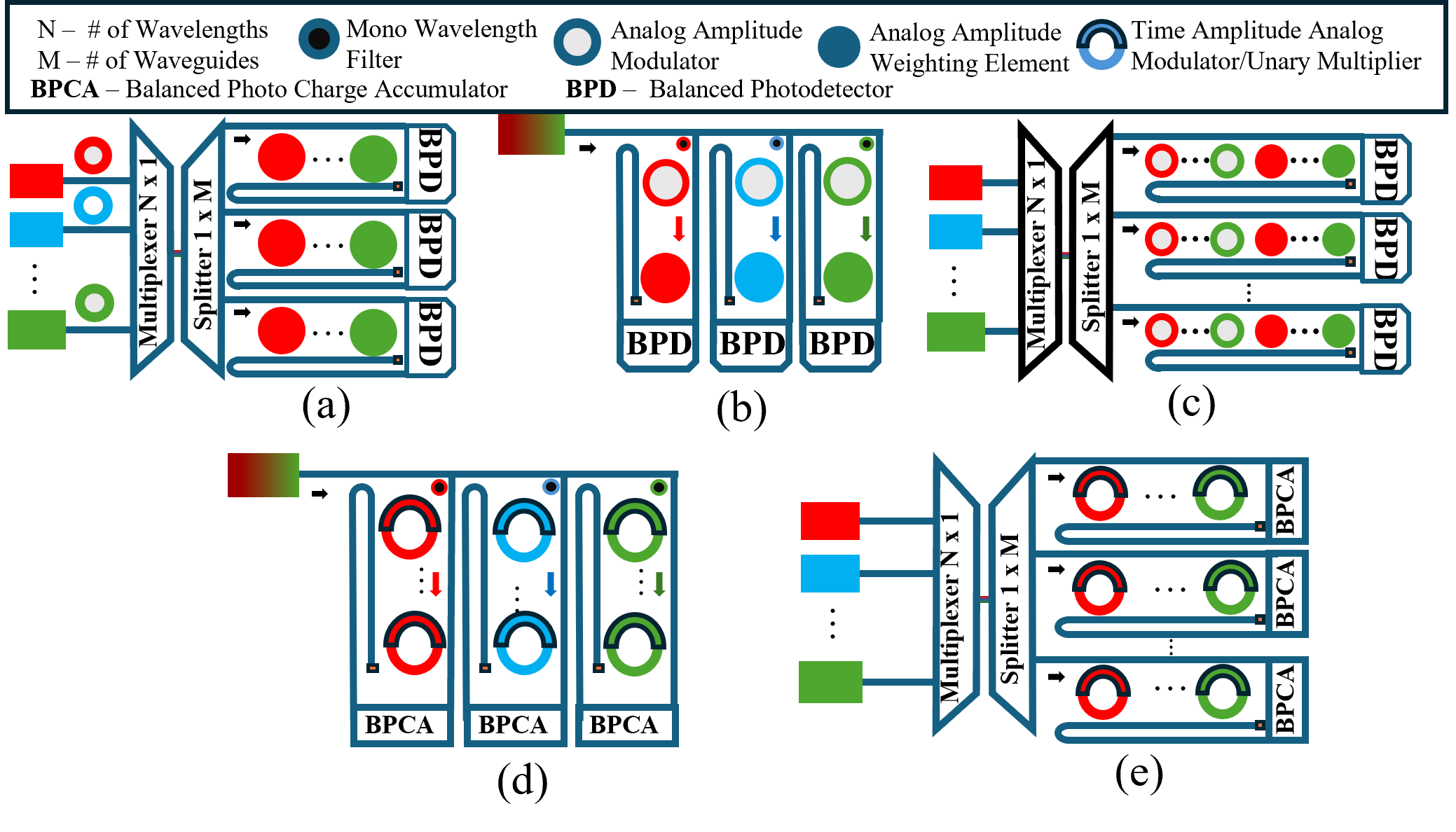}
\caption{(a) Amplitude analog MAW core,
(b) amplitude analog MWA core,
(c) amplitude analog AMW core,
(d) time analog homodyne MWA core, can be operated as unary homodyne core as well,
(e) unary heterodyne AMW core. N and M, respectively, represent the numbers of multiplexed wavelengths and waveguides per core. M$\times$N defines achievable parallel MACs per core.}
\label{fig:1}
\end{figure}

\section{Scalability Results}

We evaluate achievable number of spatially parallel multiply-accumulate (MAC) operations for every considered core type, which is defined as M$\times$N (see Fig. 1). Table~\ref{tab:total_eff_macs} shows that both organization and encoding strongly affect scalability. Among the analog cores, MWA is the best performer at every data rate, reaching 3{,}984 MACs at 1 GS/s versus 1{,}849 for MAW and 1{,}296 for AMW. This indicates that moving aggregation to the end of the signal path reduces pre-detection loss and improves usable parallelism.

The analog designs are also highly sensitive to data rate. MAW drops from 1{,}849 to 225 MACs as the rate increases from 1 GS/s to 10 GS/s, while AMW falls from 1{,}296 to 144 MACs and HEANA from 6{,}889 to 900 MACs. This trend reflects the tighter optical power budget at higher bandwidths, which makes amplitude-encoded analog signaling increasingly difficult to sustain.

In contrast, the unary designs are rate-independent in this table. SCONNA stays fixed at 15{,}840 MACs, and ASTRA stays fixed at 25{,}600 MACs across all three data rates. This confirms that unary encoding decouples precision from analog amplitude levels and preserves spatial MAC throughput under changing bandwidth.

The difference between SCONNA and ASTRA highlights the benefit of homodyne superposition. Both use unary encoding, but ASTRA achieves the highest total MAC count because homodyne accumulation avoids the wavelength-spacing and crosstalk limits of incoherent wavelength multiplexing. Overall, the table suggests three takeaways: MWA is the strongest analog organization, unary encoding is the most robust against rate increases, and homodyne superposition offers the clearest path to the highest spatial MAC count.

\begin{table}[t]
\centering
\caption{Total Spatial MACs Across Photonic Tensor Core Configurations and Data Rates}
\label{tab:total_eff_macs}
\renewcommand{\arraystretch}{1.2}

\begin{tabular}{|>{\raggedright\arraybackslash}p{3.3cm}|c|c|c|}
\hline
\textbf{Configuration} & \textbf{1 GS/s} & \textbf{5 GS/s} & \textbf{10 GS/s} \\ \hline

Analog MAW (HOLYLIGHT) & 1,849 & 441 & 225 \\ \hline
Analog AMW (DEAP-CNN) & 1,296 & 289 & 144 \\ \hline
Analog MWA (SPOGA) (M=16 fixed) & 3,984 & 2,992 & 2,544 \\ \hline
Unary AMW (@8bits) (SCONNA) (N=90 fixed) & 15,840 & 15,840 & 15,840 \\ \hline
Time Analog MWA (HEANA) & 6889 & 1764 & 900 \\ \hline
Unary Homodyne MWA (ASTRA) (N=25 fixed) & 25,600 & 25,600 & 25,600 \\ \hline

\end{tabular}
\end{table}

\section{Conclusion}

This work compared five realizable photonic microring tensor-core configurations across organization, encoding, and superposition mode. The results show that MWA is the strongest analog organization because it minimizes pre-detection loss, while unary encoding is the most effective way to keep spatial MAC count stable as data rate increases. Homodyne superposition further improves scalability by removing wavelength-spacing and crosstalk limits, which gives ASTRA the highest total spatial MAC count in the table. Overall, future photonic tensor cores should favor unary or stochastic encoding, and homodyne accumulation when maximum parallelism is the primary goal.




\end{document}